\documentclass[draft]{agujournal}

\journalname{Space Weather}

\usepackage{bm}
\usepackage{amsmath}
\usepackage{subfigure}
\usepackage[colorlinks=true, pdfstartview=FitV, linkcolor=black, citecolor= black, urlcolor= black]{hyperref}
\usepackage{overcite}
\usepackage{footnpag}			      	% make footnote symbols restart on each page
\usepackage{amssymb}
\usepackage{algorithm,algorithmic}
\usepackage{multirow,bm}
\usepackage{xcolor}
\usepackage[normalem]{ulem}
 \usepackage{subfigmat}% matrices of similar subfigures, aka small mulitples

\bibpunct{(}{)}{;}{a}{}{,}

\begin{document}

\title{Benchmarking Forecasting Models for Space Weather Drivers}

%% ------------------------------------------------------------------------ %%
%
%  AUTHORS AND AFFILIATIONS
%
%% ------------------------------------------------------------------------ %%

 \authors{Richard J. Licata\affil{1}\thanks{1306 Evansdale Drive, Morgantown, West Virginia 26506-6106.}, W. Kent Tobiska\affil{2}, and Piyush M. Mehta\affil{1}}

\affiliation{1}{Department of Mechanical and Aerospace Engineering,
West Virginia University, Morgantown, West Virginia, USA.}
\affiliation{2}{Space Environment Technologies, Pacific Palisades, California, USA.}

%% Corresponding Author
%(include name and email addresses of the corresponding author.  More
%than one corresponding author is allowed in this Word file and for
%publication; but only one corresponding author is allowed in our
%editorial system.)  

\correspondingauthor{Richard J. Licata}{rjlicata@mix.wvu.edu}

%  List up to three key points (at least one is required)
%  Key Points summarize the main points and conclusions of the article
%  Each must be 100 characters or less with no special characters or punctuation 

\begin{keypoints}
\item Four solar ($F_{10.7}$, $S_{10.7}$, $M_{10.7}$, and $Y_{10.7}$) and two geomagnetic ($a_p$ and $Dst$) driver indices used as inputs by the operational HASDM system.
\item Temporal statistics using six years of historical data set for driver forecasts.
\item Baseline for future developments within the community.

\end{keypoints}

% \maketitle{} 		

\begin{abstract}
Space weather indices are commonly used to drive operational forecasts of various geospace systems, including the thermosphere for mass density and satellite drag. The drivers serve as proxies for various processes that cause energy flow and deposition in the geospace system. Forecasts of neutral mass density is a major uncertainty in operational orbit prediction and collision avoidance for objects in low earth orbit (LEO). For the strongly driven system, accuracy of space weather driver forecasts is crucial for operations. The High Accuracy Satellite Drag Model (HASDM) currently employed by the United States Air Force in an operational environment is driven by four (4) solar and two (2) geomagnetic proxies. Space Environment Technologies (SET) is contracted by the space command to provide forecasts for the drivers. This work performs a comprehensive assessment for the performance of the driver forecast models. The goal is to provide a benchmark for future improvements of the forecast models. Using an archived data set spanning six (6) years and 15,000 forecasts across solar cycle 24, we quantify the temporal statistics of the model performance. 

% Forecasting thermospheric neutral mass density is a major challenge due the thermosphere's highly driven nature and the impact of complex coupling within the geospace environment. Even if a density model itself is perfect, the only way to accurately forecast density would be with perfect driver forecasts. To improve satellite orbit determination, the accuracy of density model driver forecasts must be improved. The current state-of-the-art drag model is the United States Air Force (USAF) High Accuracy Satellite Drag Model (HASDM). This model uses various algorithms from Space Environment Technologies (SET) that provide six-day forecasts for many solar and geomagnetic indices which are drivers for the JB2008 density model. Using over 15,000 forecasts, this paper serves to benchmark these algorithms by comparing them to the issued values across a variety of solar and geomagnetic conditions. 
\end{abstract}

\section{Introduction}\label{sec:intro}

Accurately quantifying mass density in the thermosphere remains a predicament for the community. The difficulty stems from the highly dynamic nature of the thermosphere, an environment driven by a number of factors ranging from solar extreme ultraviolet (EUV) and geomagnetic heating to gravity waves in the lower atmosphere. Emmert \citeyearpar{Emmert07} provides a thorough overview of the physical drivers and their effects on thermospheric density. Current capabilities limit our ability to predict satellites' trajectories with precision in an operational setting. During large solar and geomagnetic storms, operators struggle to locate many resident space objects, let alone have the means to predict their orbits \citep{Berger}. Many resources in the United States and abroad are devoted to tracking satellites and determining their orbits in order to protect humans and other assets in space.

HASDM \citep{HASDM} is an assimilative empirical model that uses a large batch of calibration satellites to make corrections to a density nowcast from the Jacchia-Bowman 2008 (JB2008) model \citep{JB2008,JB2008_12}.  The resulting density data cube is propagated in time using driver forecasts supplied to JB2008. The forecasts are deterministic in nature.

The JB2008 models neutral density in the thermosphere using global exospheric temperature equations that leverage four solar indices to simulate thermosphere heating from different sources of solar energy \citep{Indices,JB2008}. The $F_{10.7}$ proxy has a strong correlation to solar extreme ultraviolet (EUV) irradiance which has led to its long-time use as measure of solar EUV energy. $S_{10.7}$ is an index indicative of activity of the integrated 26-34 nm bandpass solar chromospheric EUV emission, which penetrates to the middle thermosphere and is absorbed by atomic oxygen. The $M_{10.7}$ proxy is used as a measure of far ultraviolet (FUV) photospheric 160 nm Schumann-Runge Continuum emissions, which penetrate to the lower thermosphere and cause molecular oxygen dissociation. The fourth solar index is $Y_{10.7}$ which is a composite of $X_{b10}$ and Lyman-alpha. This serves as a composite measure of solar coronal 0.1-0.8 nm X-ray emissions and 121.6 nm Lyman-alpha, both of which penetrate to the mesosphere and participate in water chemistry. In order to forecast these indices/proxies, SET uses a linear predictive algorithm that captures persistence and recurrence \citep{Indices}.

%\textcolor{red}{provide a brief intro to HASDM segwaying into the drivers} The most influential drivers that will be focused on are EUV heating and the effects of energy dissipation from energetic particles entering from the magnetosphere. The JB2008 model uses $F_{10.7}$, $S_{10.7}$, $M_{10.7}$, and $Y_{10.7}$ to compute the direct effects of solar irradiance. In order to forecast these indices, $F_{10.7}$, $S_{10.7}$, and $M_{10.7}$ are centered and smoothed using a moving boxcar method, creating a running 81-day set of values. $Y_{10.7}$ requires additional processing to combine $X_{10}$ and Lyman-$\alpha$, weighted with a normalized $F_{10.7}$ function \citep{Indices}.

To capture the impact of geomagnetic activity, the model uses a synthesis of $ap$ and $Dst$ indices. The $ap$ index is a measure of global geomagnetic activity derived from twelve observatories that fall between 48$^{\circ}$ N and 63$^{\circ}$ S in latitude \citep{Vallado}. The utilization of $ap$ during quiet geomagnetic conditions results in low density errors, but $Dst$ proves to be a more effective driver during storm times \citep{JB2008}. $Dst$ is an index that represents the strength of the storm-time ring current in the inner-magnetosphere \citep{Indices}. Its forecast is generated using SET's \textit{Anemomilos} algorithm, which provides a forecast with maximum prediction window of six days \citep{Anemomilos} using a data driven deterministic algorithm. For further details on all of the JB2008 drivers, see Tobiska et al. \citeyearpar{Indices} and \citet{ISO14222}.

In contrast to the other indices, $ap$ does not have an algorithm or model to provide forecasts to the JB2008 model. The three-hourly $ap$ forecasts are actually interpolated values from the National Oceanic and Atmospheric Administration (NOAA) Space Weather Prediction Center's (SWPC) $Kp$ forecasts (\url{https://www.swpc.noaa.gov/products/planetary-k-index}). Additionally, they are generated from an ensemble of individual human forecasters' predictions informed by model output [University of Michigan's Geospace Model since 2017] \citep{Steenburgh,Singer,Geospace}. This forecast only extends three days, so the value is set to zero for the last three days of each window. Even though SWPC only recently switched to using the Geospace Model, this data represents the official NOAA SWPC forecast and we use it as such.

%After the JB2008 model provides a density nowcast, Air Force's HASDM model makes corrections using density estimates derived from a large batch of calibration satellite trajectories \citep{HASDM}. The resulting density array is propagated out in time using the driver forecasts and JB2008. The ensuing density predictions show to be more robust with respect to varying solar and geomagnetic conditions \citep{Doornbos}. This all relies on the accuracy of driver forecasts provided by SET's algorithms and NOAA SWPC forecasters. Knowledge on driver uncertainty is critical in performing uncertainty quantification, especially on satellite trajectories as shown by Licata et al. \citeyearpar{Licata}. Results from this work are displayed in Figure \ref{f:prob}.

%\textcolor{red}{After the JB2008 model forecasts density, Air Force's HASDM model makes corrections using density estimates derived from a large batch of calibration satellite trajectories \citep{HASDM}. The resulting density shows to be more robust with respect to varying solar and geomagnetic conditions \citep{Doornbos}. This all originates from the driver forecasts provided by SET's algorithms and NOAA SWPC forecasters. Knowledge on driver uncertainty is critical in performing uncertainty quantification, especially on satellite trajectories as shown by Licata et al. \citeyearpar{Licata}. Results from this work are displayed in Figure \ref{f:prob}.} \textcolor{blue}{I am not sure about this, Kent can you please review and confirm?}

Errors in the space weather driver forecasts cause errors in the resulting densities, therefore impairing satellite conjunction analyses. Bussy-Virat et al. \citeyearpar{Bussy-Virat} recently performed a study to show the effects on driver uncertainty on the probability of collision between two space objects. A similar study was performed more recently by Licata et al. \citeyearpar{Licata} incorporating additional forecasts and further conditioning distributions.

%\textcolor{red}{I think we need to make the sampling for $F_{10.7}$ more realistic. Bussy-Virat kinda did this correctly (I do not agree with what they did for ap). $F_{10.7}$ will not have crazy variations. When the initial value is negative or positive, the likelihood of the future values keeping the same trend is high. We need to incorporate that knowledge somehow in the sampling.}

\begin{figure}[htb!]
	\centering
	\small
	\includegraphics[width=\textwidth]{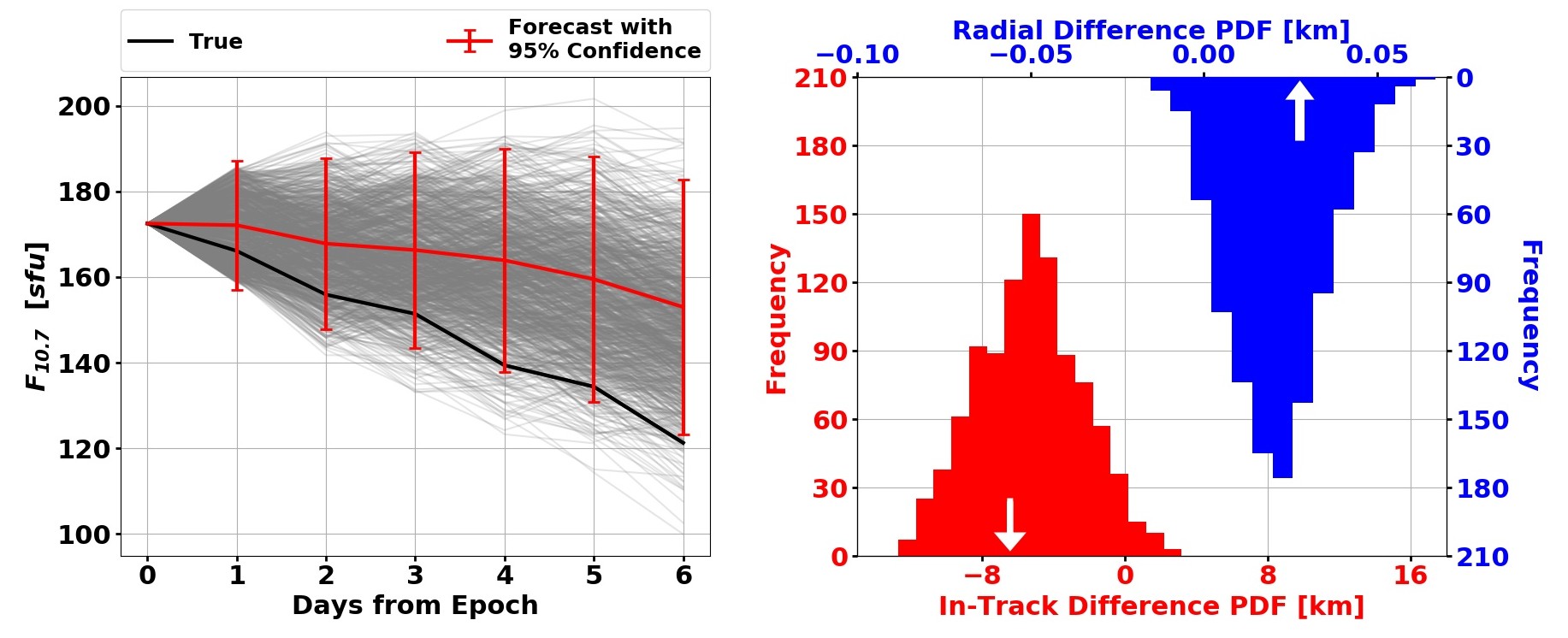}
	\caption{(left) Deterministic and probabilistic $F_{10.7}$ forecasts in addition to the true variation during the time period. (right) Satellite position distributions relative to the true position after encountering six days of probabilistic densities resulting from the corresponding $F_{10.7}$ fluctuations. White arrows represent position using deterministic $F_{10.7}$ values.}
	\label{f:prob}
\end{figure}

The probabilistic $F_{10.7}$ forecasts in Figure \ref{f:prob} were generated using the statistical measures identified in the current study. There was a constraint of the maximum change in the driver ($dF_{10.7}$) from one time-step to the next. This limiting factor was chosen through further statistical analyses. Each driver forecast was input to a quasi-physical model of the mass density built using recurrent neural network to forecast a resulting 3D density grid that would be used in orbit propagation \citep{MehtaROM,AGU,AMS}. The satellite position distributions give light to the need for probabilistic approaches in determining satellites' orbits. In this fairly quiet case, there was a $\sim$6.4 km position error with deterministic approaches. Probabilistic forecasting allows for the true position to be captured through the analysis. Here, the mean probabilistic position was more accurate than the deterministic position. Figure \ref{f:prob} is a derivative of this work.

We expand upon the work of Bussy-Virat et al. \citeyearpar{Bussy-Virat} by using (i) all solar and geomagnetic drivers that are used in operations, (ii) a large historical data set covering a period of six (6) years, (iii) an extended forecast window of up to six (6) days, and (iv) the initial driver values to characterize model performance as a function of the solar and geomagnetic activity.

% Bussy-Virat et al. \citeyearpar{Bussy-Virat} had characterized three-day temporal forecast uncertainty using two years of forecasts and historical data for $F_{10.7}$ and $ap$ from SWPC. To expand upon that work, this paper incorporates four additional model drivers, employs six years of forecasts, extends the forecast window to six days, and focuses on system variability from activity level.

The outline for the current paper is as follows: the following section introduces the techniques and thresholds to bin solar and geomagnetic drivers. This is done separately between the domains and presents distinct methods. Next, the resulting uncertainty figures are presented and discussed followed by the conclusion.

%%%%%%%%%%%%%%%%%%%%%%%%%%%%%%%%%%%%%%%%%%%%%%%%%%%%%%%%%%%%%%%%%%%%%%%%%%%%%%%%%%%%%%%%%%%%%%%%%%%%%%%%%%%%%%
\section{Methodology}\label{sec:method}
The SET algorithms produce files every three hours generating updated six-day forecasts for solar and geomagnetic indices. These forecasts have a temporal resolution of three hours. In addition, they archive the observed values for each time step. To conduct this analysis, forecasts from October 2012 through the end of 2018 were used with the exception of some missing/corrupted forecasts. In total, there were over 15,000 files to leverage for this study.

In order to effectively examine the solar and geomagnetic indices in comparable terms, a consistent approach had to be determined. To provide the clearest possible representation for all indices, different methods are used for solar indices and geomagnetic indices but kept consistent within the domains. Each index was split into separate sub-populations depending on the initial forecasted value. Populations that ended up with fewer than 100 forecasts are not shown, because there is insufficient data to draw statistical conclusions.

\subsection{Solar Indices}
\begin{figure}[htb!]
	\centering
	\small
	\includegraphics[width=\textwidth]{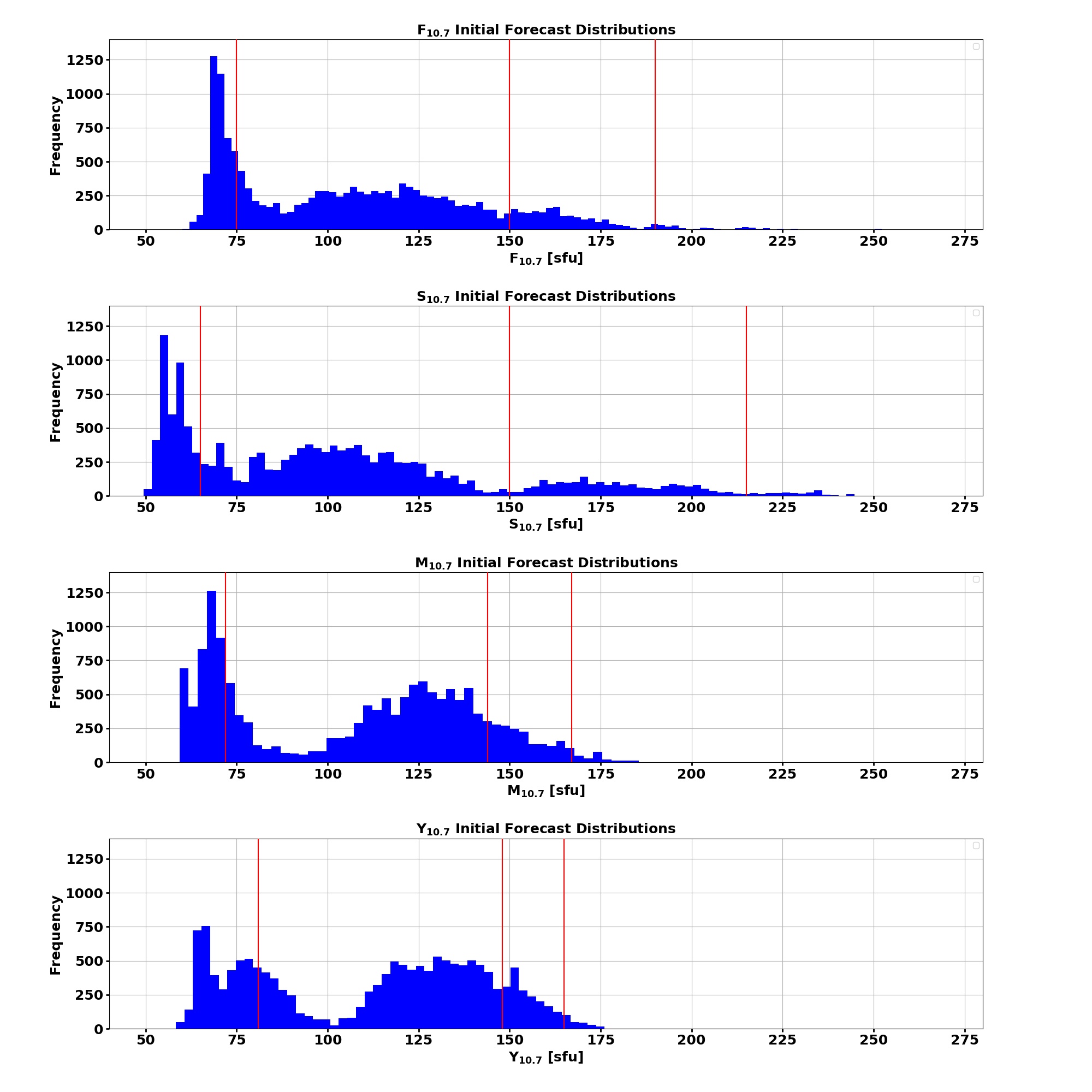}
	\caption{Distributions of initially forecasted values for each solar index with partitions shown in red.}
	\label{f:Distr}
\end{figure}

The task of generating statistical results for the four (4) solar indices investigated ($F_{10.7}$, $S_{10.7}$, $M_{10.7}$, and $Y_{10.7}$) was relatively straightforward. The forecasts are generated using SET's \textit{SOLAR2000} algorithm \citep{TOBISKA2000,TOBISKA2008}. The thresholds to assess activity level were determined by the experience of previous work combined with a supplementary statistical analysis. Figure \ref{f:Distr} depicts how the solar indices are distributed based on the initially forecasted value.

The thresholds for $F_{10.7}$ had been previously specified and used in previous work \citep{PMMthesis,Licata}. Using these partitions on the 15,000+ forecasts resulted in a distinct number of individual $F_{10.7}$ forecasts for each activity level. This was used to classify the remaining solar indices, with the absence of a natural partition. A natural partition within a distribution is seen at 150 sfu for $S_{10.7}$. This was chosen for the particular threshold as it did not greatly disrupt the number of forecasts in the adjacent activity levels. The four levels of solar activity are defined in Table \ref{t:Solar Level}.

\begin{table}[htb!]
	\fontsize{10}{10}\selectfont
    \caption{Activity level thresholds for the four solar indices.}
   \label{t:Solar Level}
        \centering 
   \begin{tabular}{r | c | c} % Column formatting, 
      \hline 
            \multirow{4}{*}{\textbf{\textit{$F_{10.7}$}}} & Low & $F_{10.7}$ $\leq$ $75$ \textrm{ sfu}\\
            & Moderate & $75 <$ $F_{10.7}$ $\leq$ $150$ \textrm{ sfu}\\
            & Elevated & $150 <$ $F_{10.7}$ $\leq$ $190$ \textrm{ sfu}\\
            & High & $F_{10.7}$ $>$ $190$ \textrm{ sfu}\\ \hline
            \multirow{4}{*}{\textbf{\textit{$S_{10.7}$}}} & Low & $S_{10.7}$ $\leq$ $65$\\
            & Moderate & $65 <$ $S_{10.7}$ $\leq$ $150$\\
            & Elevated & $150 <$ $S_{10.7}$ $\leq$ $215$\\
            & High & $S_{10.7}$ $>$ $215$ \\ \hline
            \multirow{4}{*}{\textbf{\textit{$M_{10.7}$}}} & Low & $M_{10.7}$ $\leq$ $72$\\
            & Moderate & $72 <$ $MS_{10.7}$ $\leq$ $144$\\
            & Elevated & $144 <$ $M{10.7}$ $\leq$ $167$\\
            & High & $M_{10.7}$ $>$ $167$ \\ \hline
            \multirow{4}{*}{\textbf{\textit{$Y_{10.7}$}}} & Low & $Y_{10.7}$ $\leq$ $81$\\
            & Moderate & $81 <$ $Y_{10.7}$ $\leq$ $148$\\
            & Elevated & $148 <$ $Y_{10.7}$ $\leq$ $165$\\
            & High & $Y_{10.7}$ $>$ $165$ \\
      \hline
   \end{tabular}
\end{table}

With each index's forecast appropriately divided on initial forecasted value, uncertainty distributions could be generated with respect to time from epoch. The uncertainty for the solar indices is defined as the error with respect to the issued value, normalized by the issued value. It is important to note that all errors shown (for both solar and geomagnetic indices) have a consistent sign convention. Positive percentages represent a forecasted value that was \textbf{more positive} than the issued value. All of the solar indices are updated daily, there are twenty-four distributions for each (four magnitude-based and six temporal partitions).

\subsection{Geomagnetic Indices}
The analysis of the two geomagnetic indices, $ap$ and $Dst$, was more intricate. Not only are the uncertainties functions of their magnitudes and time from epoch, they vary with solar activity level. To analyze $ap$, three geomagnetic activity levels were chosen: low, moderate and active. In analyzing $Dst$, six geomagnetic activity levels were chosen and are consistent with the NOAA G-scale as operationally applied by SET. Table \ref{t:Geomagnetic Level} states the thresholds for $ap$ and $Dst$.

\begin{table}[htb!]
	\fontsize{10}{10}\selectfont
    \caption{Bin thresholds for geomagnetic activity, $ap$ and $Dst$.}
   \label{t:Geomagnetic Level}
        \centering 
   \begin{tabular}{r | c | c} % Column formatting, 
      \hline 
            \multirow{3}{*}{\textbf{\textit{$ap$}}} & Low & $ap$ $\leq$ $10$\\
         & Moderate & $10 <$ $ap$ $\leq$ $50$\\
            & Active & $ap$ $>$ $50$ \\ \hline
            \multirow{6}{*}{\textbf{\textit{$Dst$}}} & G0 & $Dst$ $\geq$ $-30$\\
         & G1 & $-30$ $>$ $Dst$ $\geq$ $-50$\\
            & G2 & $-50$ $>$ $Dst$ $\geq$ $-90$\\
            & G3 & $-90$ $>$ $Dst$ $\geq$ $-130$\\
            & G4 & $-130$ $>$ $Dst$ $\geq$ $-350$\\
            & G5 & $Dst$ $\leq$ $-350$\\
      \hline
   \end{tabular}
\end{table}

To allocate the geomagnetic forecasts, the largest value in the forecast for $ap$ and the most negative value for $Dst$ is the controlling factor. In addition, the forecast is classified by the initial forecasted $F_{10.7}$ value. Since the distributions have a finer temporal resolution and a solar dependency, there are 1,152 distributions for $ap$ and $Dst$.

It becomes difficult to generate a standard percent error normalized by the issued value, because the issued value can be small or even zero. Therefore, another method had to be chosen to provide a similar comparison. Instead of normalizing errors by the issued value, they are normalized by the long-term mean value of the index. Therefore, an error of $-200\%$ for $Dst$ signifies an error twice the magnitude of the long-term mean $Dst$, and the prediction was more-negative than the issued value. The long-term mean values for \textit{ap} and \textit{Dst} are 9.2 \textit{2nT} and -8.8 \textit{nT}, respectively.

%%%%%%%%%%%%%%%%%%%%%%%%%%%%%%%%%%%%%%%%%%%%%%%%%%%%%%%%%%%%%%%%%%%%%%%%%%%%%%%%%%%%%%%%%%%%%%%%%%%%%%%%%%%%%%
\section{Results}\label{sec:results}
\begin{figure}[htb!]
	\centering
	\small
	\includegraphics[width=0.85\textwidth]{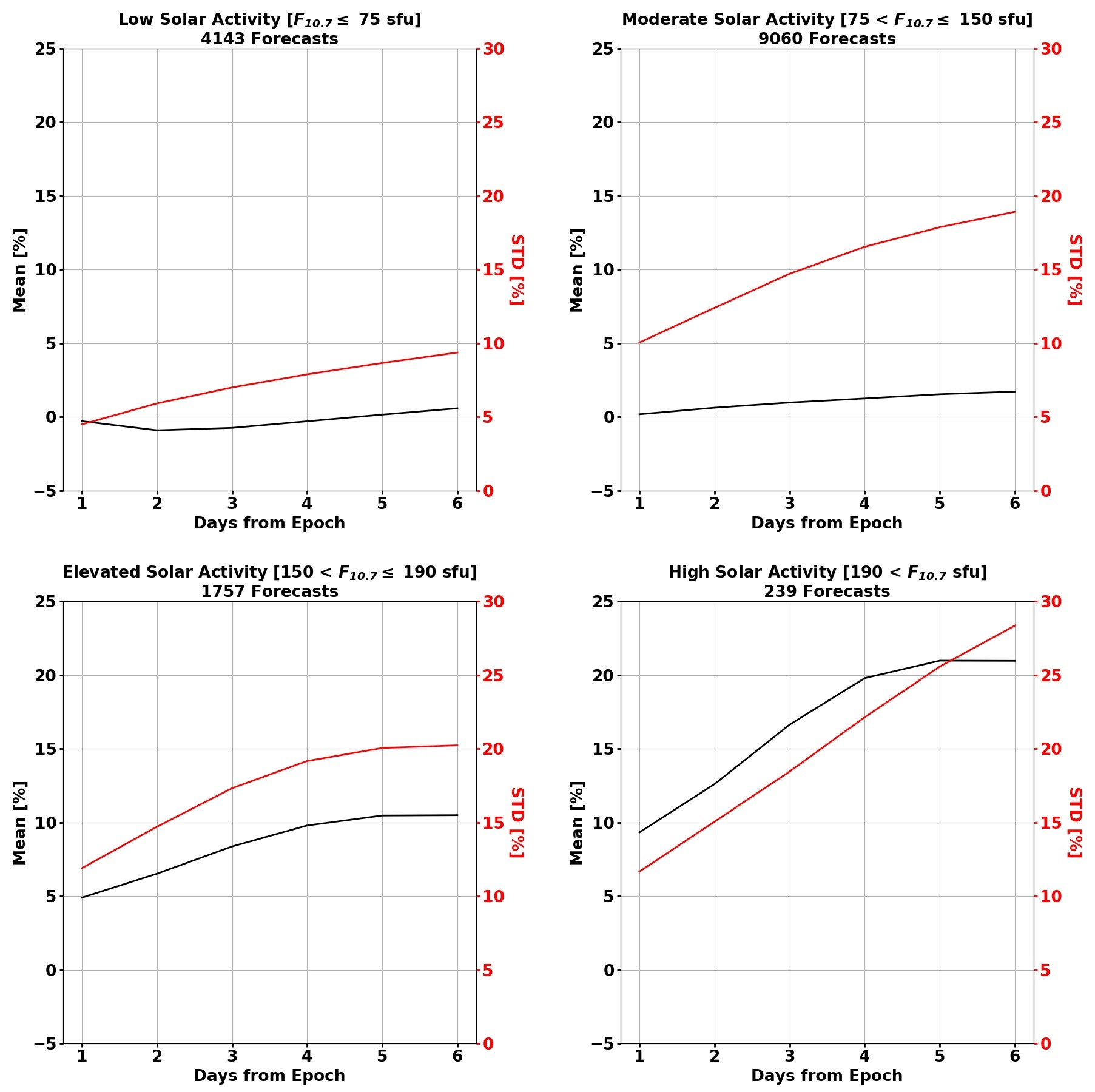}
	\caption{$F_{10.7}$ algorithm performance across four levels of solar activity.}
	\label{f:F10}
\end{figure}

In the resulting uncertainty figures, the mean and standard deviation of forecast error (as a function of time from current epoch) are presented for each activity level. This way, biases can be identified and the algorithm's temporal uncertainty can be determined. Figure \ref{f:F10} shows the performance of the $F_{10.7}$ forecast algorithm. 

At low and moderate levels of solar activity, the $F_{10.7}$ algorithm is fairly unbiased. It is not until elevated and high solar activity that a bias accumulates, showing a tendency of over-forecasting the index. The evolution of the error's standard deviation has an expected growth with time from epoch for all activity levels, showing the uncertainty of the forecast increasing with time. The algorithm performs well when the first forecasted $F_{10.7}$ value is below 150 sfu, which accounted for approximately $87\%$ of the forecasts.

\begin{figure}[htb!]
	\centering
	\small
	\includegraphics[width=0.85\textwidth]{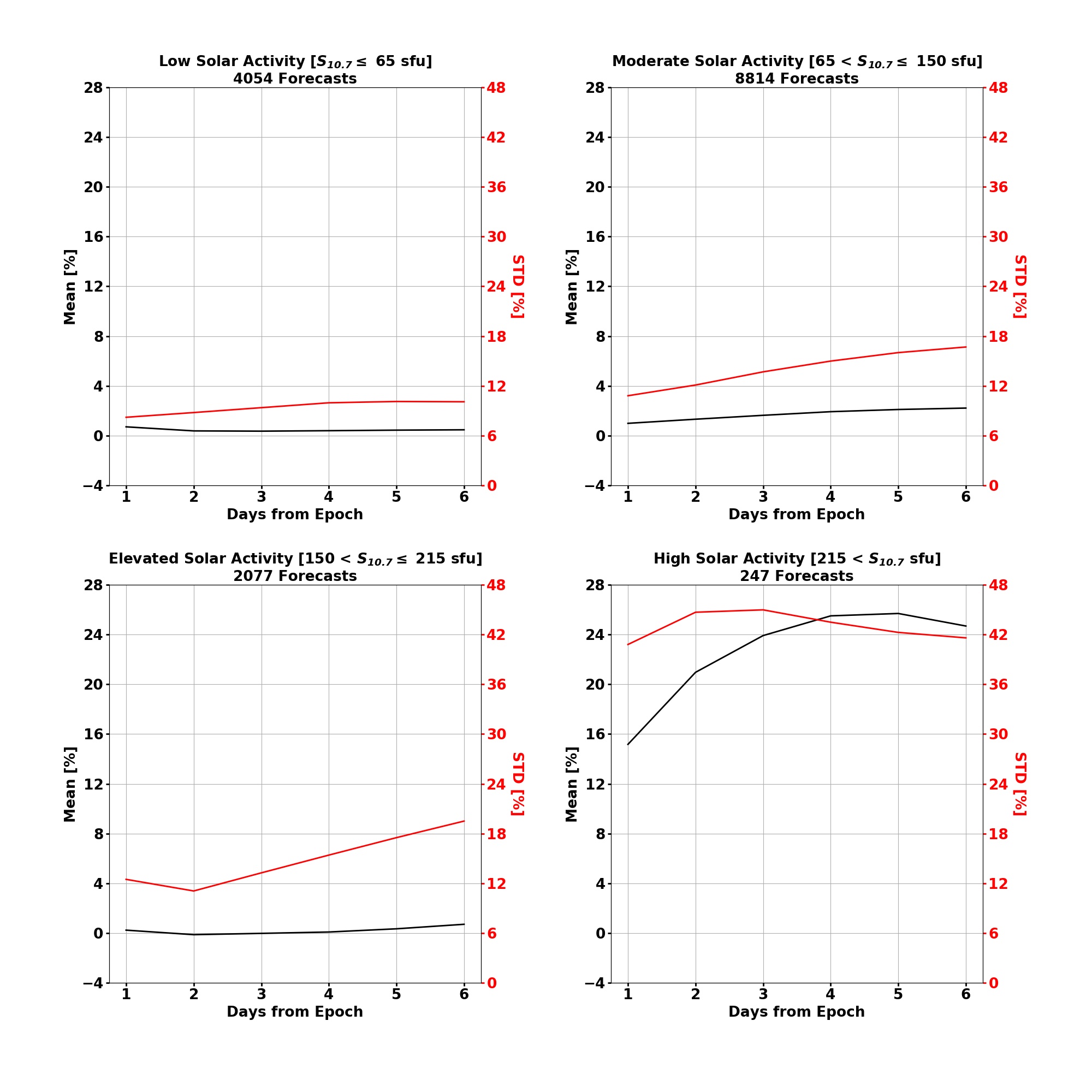}
	\caption{$S_{10.7}$ algorithm performance across four levels of solar activity.}
	\label{f:S10}
\end{figure}

Figure \ref{f:S10} provides the algorithm performance for $S_{10.7}$. There is little bias through low, moderate, and elevated activity levels (over $98\%$ of forecasts) displaying strong overall performance. The uncertainty at these activity levels is similar to $F_{10.7}$, but the performance at high solar activity is not as stable. For high solar activity, there is a dominant tendency to over-forecast in addition to a large uncertainty. The uncertainty also does not consistently grow with time.

\begin{figure}[htb!]
	\centering
	\small
	\includegraphics[width=0.85\textwidth]{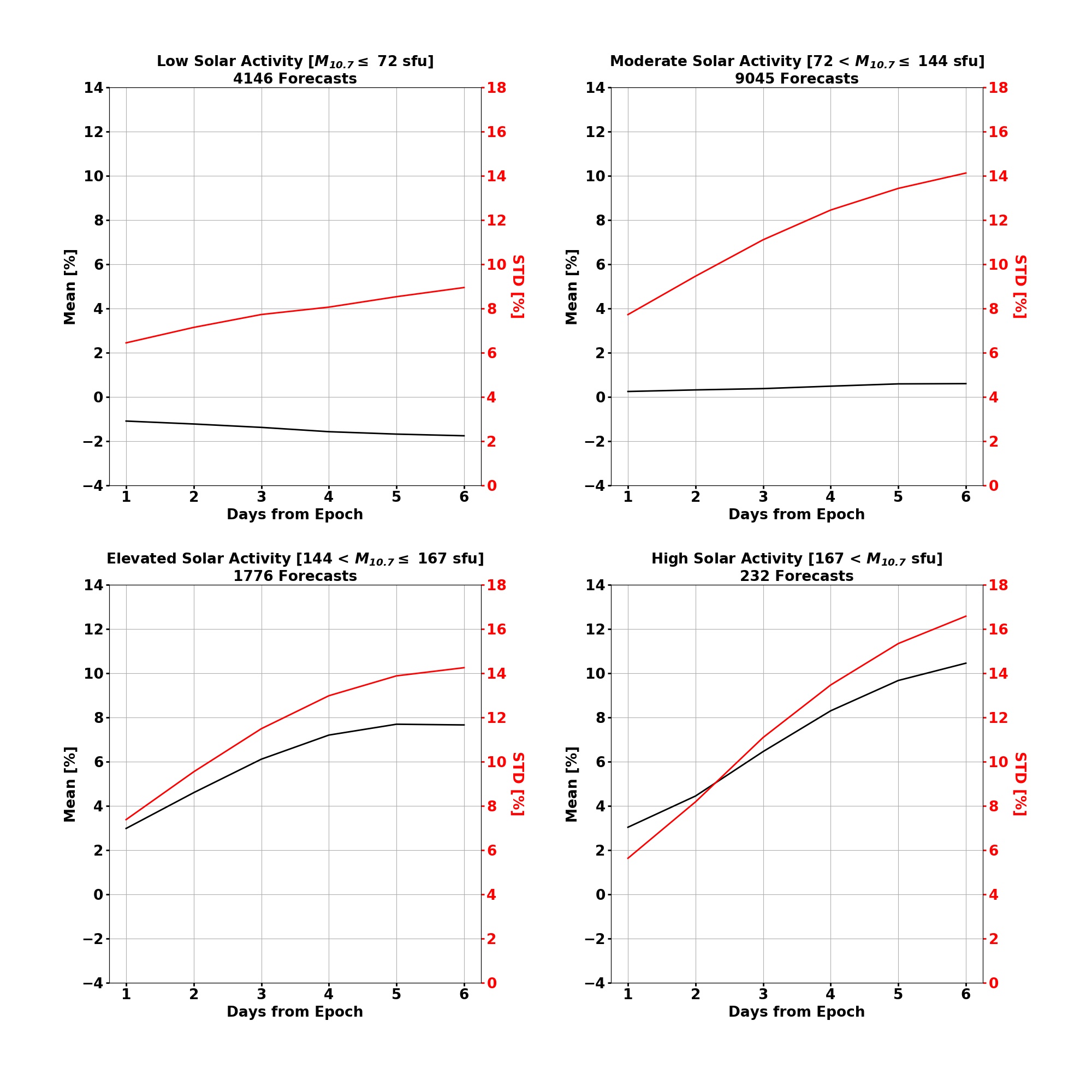}
	\caption{$M_{10.7}$ algorithm performance across four levels of solar activity.}
	\label{f:M10}
\end{figure}

The $F_{10.7}$ and $S_{10.7}$ algorithms are both vulnerable to high solar activity, but the comprehensive effectiveness is visible. The limitation during high activity is due to the volatility of the sun during solar maximum, i.e, the inability to accurately forecast flares and the lack of information from the solar East limb and solar far-side active region's growth. The algorithms for the remaining indices prove to be more robust to solar activity. The $M_{10.7}$ performance is presented in Figure \ref{f:M10}.

For $M_{10.7}$, there is a minimal bias of $\pm2\%$ for the lower two activity levels, but at low solar activity, there is a slight tendency to under-predict. At elevated and high solar activity, the bias is accumulating with time and increases in intensity. Across all levels, the uncertainty starts below $4\%$ and grows steadily with time. An interesting characteristic that contrasts the prior two indices is the lower uncertainty at high solar activity. The difference in performance is not drastic relative to the other conditions.

\begin{figure}[htb!]
	\centering
	\small
	\includegraphics[width=0.85\textwidth]{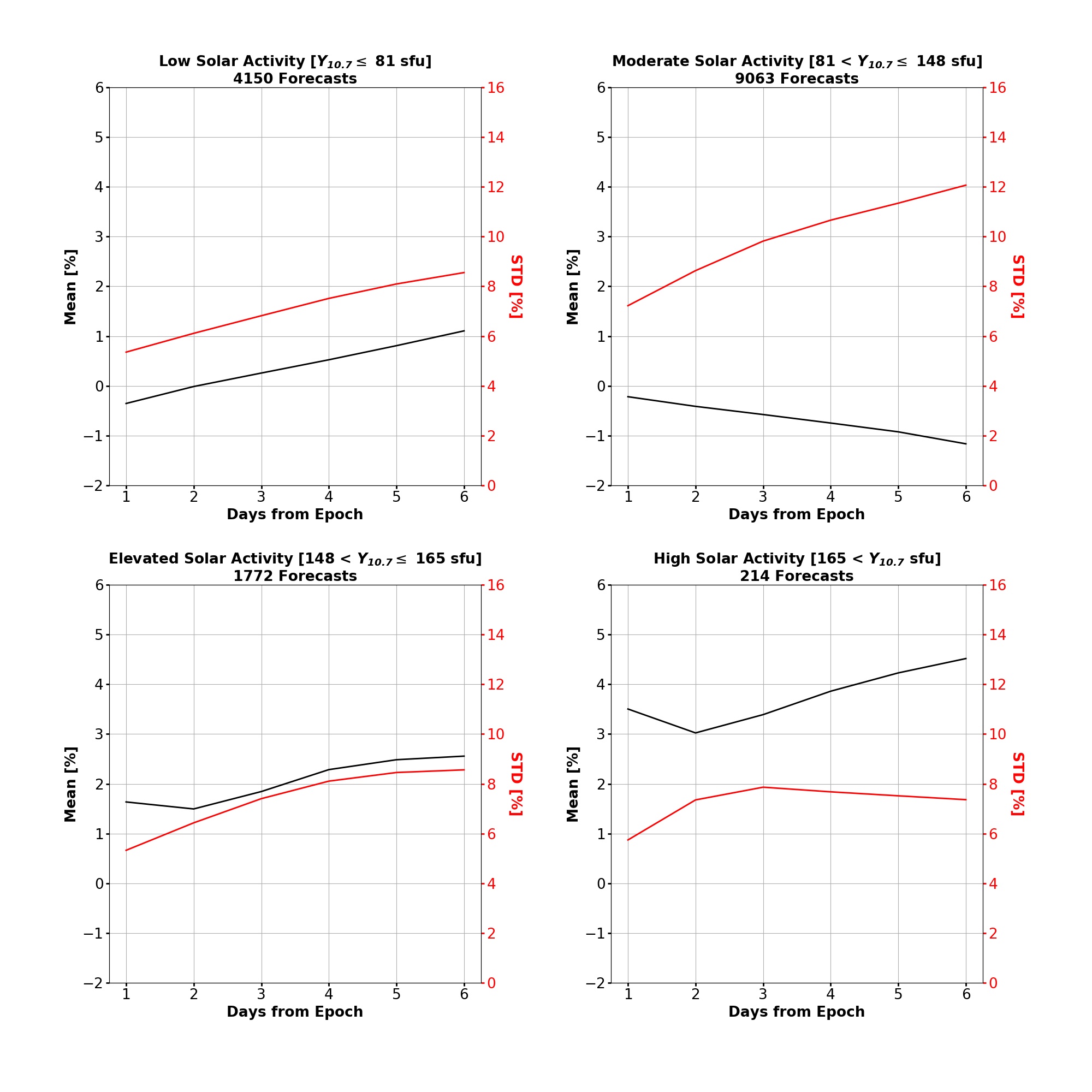}
	\caption{$Y_{10.7}$ algorithm performance across four levels of solar activity.}
	\label{f:Y10}
\end{figure}

To conclude the analysis of the solar indices, Figure \ref{f:Y10} shows the performance of the $Y_{10.7}$ algorithm. Relative to the previous three indices, the $Y_{10.7}$ algorithm is considerably robust to activity levels and has less overall uncertainty. In the first two activity levels, the bias is less than $\pm1\%$ for nearly the entire prediction window. The uncertainty grows with time for all activity levels, but its magnitude is less significant than the other indices. The bias never exceeds $5\%$ and the uncertainty $12\%$.

\begin{figure}[htb!]
	\centering
	\small
	\includegraphics[width=\textwidth]{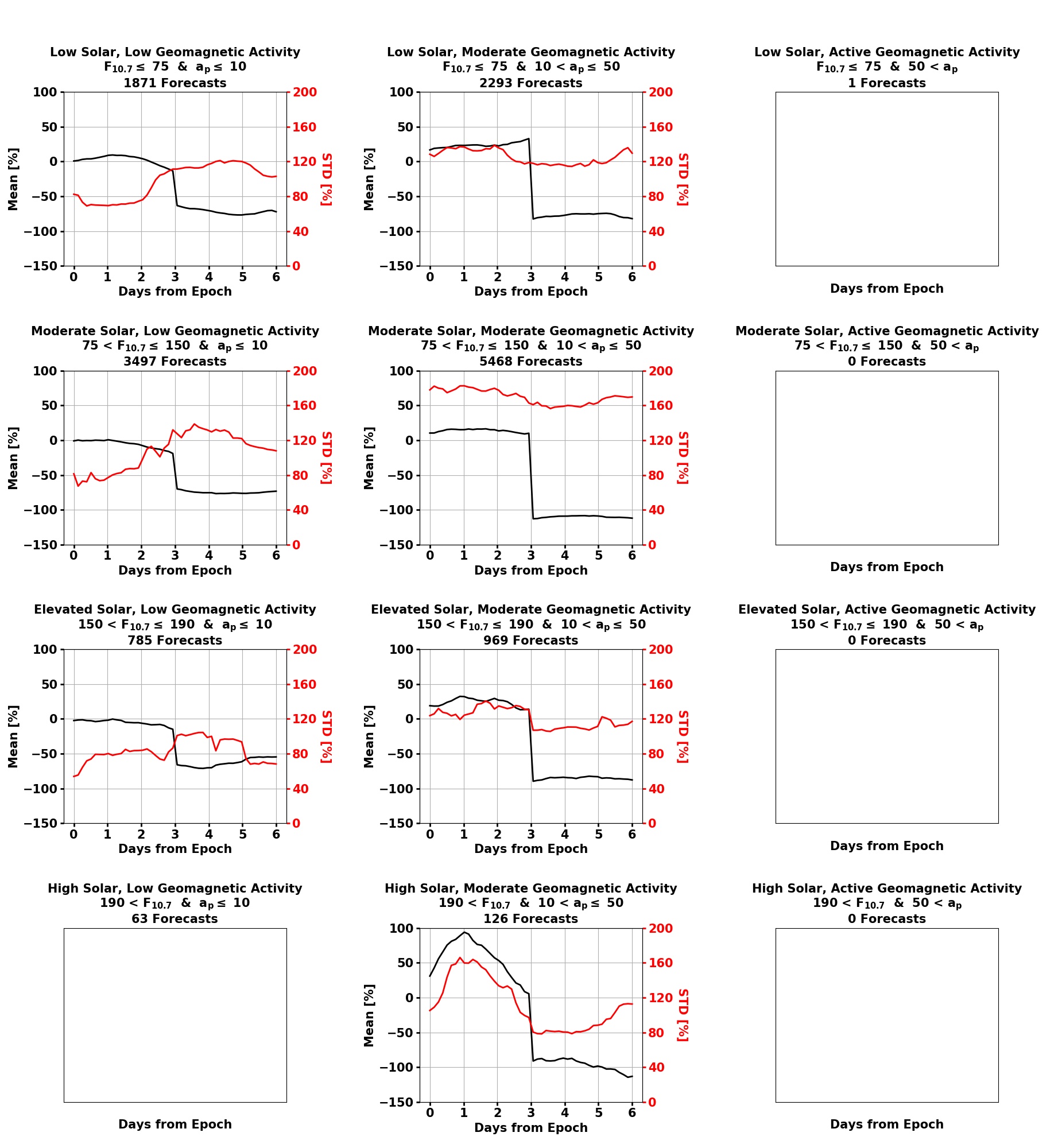}
	\caption{$ap$ forecast uncertainty for the twelve solar and geomagnetic conditions.}
	\label{f:ap}
\end{figure}

As previously stated, the geomagnetic indices were more difficult to analyze due to an increase in dependencies and a finer time resolution. Each geomagnetic index has its own set of activity levels but are both based on the previous $F_{10.7}$ thresholds. The performance of the $ap$ forecasts is shown in Figure \ref{f:ap}.

Unlike the solar indices, there are multiple conditions with insufficient data to conduct the analysis. The most distinct difference in the $ap$ forecast performance, relative to the other indices, is the discontinuity at the three-day mark. Mentioned in the Introduction, the forecasts only have a three-day prediction window. The forecasts are provided by the judgement of an array of Space Weather forecasters at NOAA SWPC with the aid of the Geospace model.

Figure \ref{f:ap} shows uncertainty results for a six-day prediction window to be consistent with the other indices, even though SET sets every $ap$ value to zero after three days. There are still interesting results in the latter three days of the forecasts across the different conditions. For example, the magnitude of under-prediction (when $ap$ is set to zero) is different for each condition as is the volatility of $ap$, shown by the standard deviation. Even so, the most important aspect of Figure \ref{f:ap} is the first three days when forecasts are provided.

During low geomagnetic activity (across all solar activity levels), there is no significant bias detected. With moderate geomagnetic activity, there is a general over-prediction that decreases over the three-day provided forecast. It shows a possible path for prediction improvement by relying on persistence when $ap$ is high at the start of the forecasts. Another key determination is shown by the right-most panels where there is only a single forecast that has a value greater than 50 $2nT$. This reflects the difficulty in quantifying the intensity of a storm, even with the aid of a physics-based model.

\begin{figure}[htb!]
	\centering
	\small
	\includegraphics[width=\textwidth]{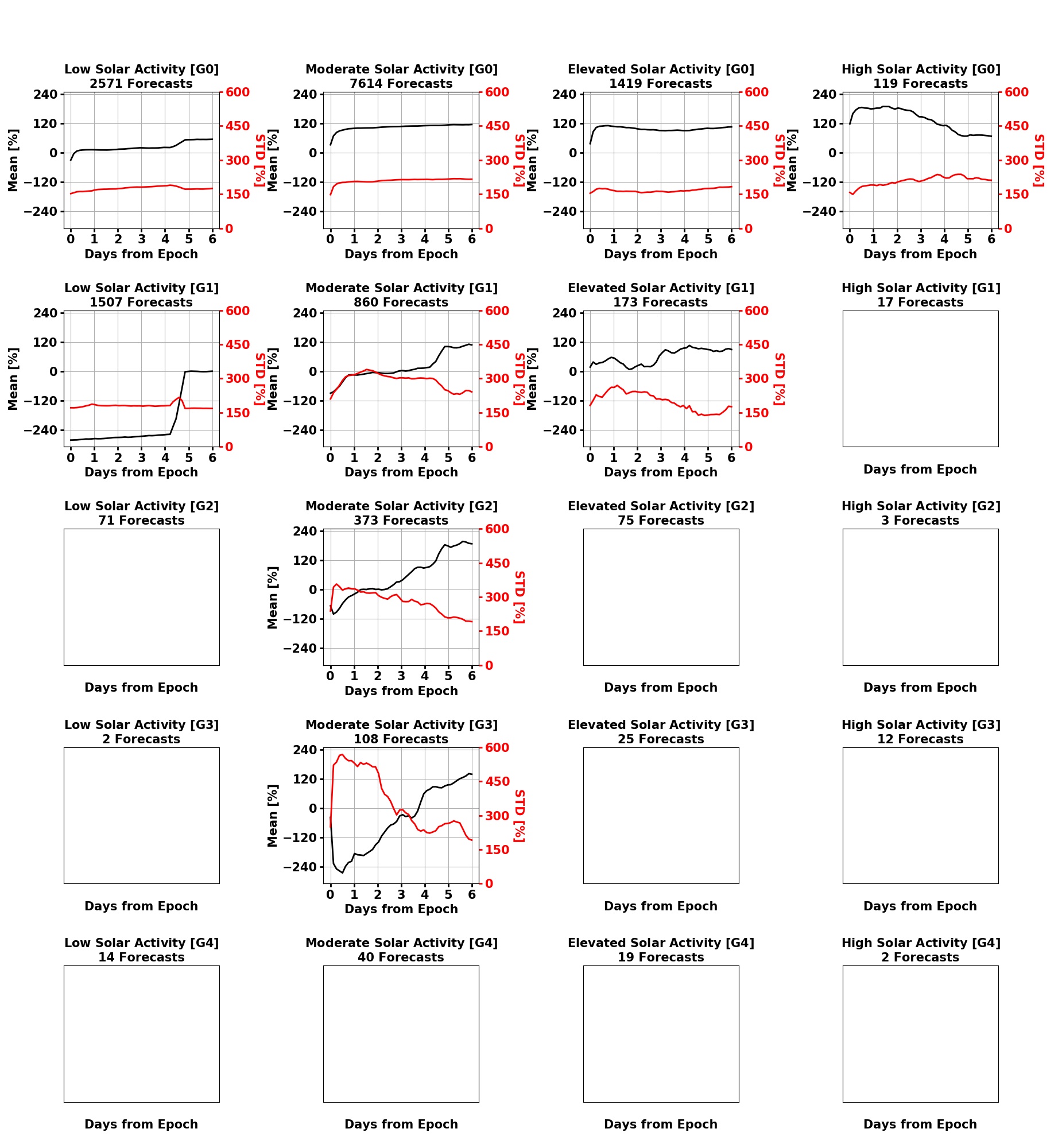}
	\caption{$Dst$ forecast uncertainty for the combined solar and geomagnetic conditions.}
	\label{f:Dst}
\end{figure}

The last algorithm analyzed is SET's \textit{Anemomilos} for $Dst$ forecasts, shown in Figure \ref{f:Dst}. The G5 row is not shown. There was only a single forecast where a G5 storm was expected. There are only $9/24$ conditions with enough forecasts to perform the analysis, but the remaining results provide insight to the strengths and weaknesses of the algorithm.

In the top-left subplot (when conditions are quiet), the forecasts remain relatively unbiased, and the uncertainty slowly increases with time. Figure \ref{f:Dst} shows a general tendency to predict $Dst$ to be more positive for nearly all G0 and G1 conditions, with the exception of G1 low solar activity conditions. In this case, the algorithm has a strong bias to expect $Dst$ to be $\sim23nT$ more negative than the issued values over the first four days of the forecast. Following the strong inclination after day four, the algorithm tends to neutralize the bias. This is interpreted as accurate prediction of $Dst$ recovery to quiet conditions. 

The bias for G1-G3, moderate solar activity conditions shows a strong temporal dependency transitioning from under to over prediction in each case. G2 moderate solar activity is a case with a peculiar trend of the uncertainty decaying with time from epoch. This is also the case for G3 moderate solar activity. However, this case has extreme and unclear results with both the bias and uncertainty changing rapidly with an inverse relationship. This behavior points to a need for improvement in these conditions. A source of the Dst variability in G0-G3 conditions is the high-speed stream (HSS) and Anemomilos does not model these events.

%%%%%%%%%%%%%%%%%%%%%%%%%%%%%%%%%%%%%%%%%%%%%%%%%%%%%%%%%%%%%%%%%%%%%%%%%%%%%%%%%%%%%%%%%%%%%%%%%%%%%%%%%%%%%%
\section{Conclusions}\label{sec:con}
The analysis of the SET algorithms used by the JB2008 and HASDM models provided clear performance capabilities for the current standard for density model driver forecasts. This work showed the many strengths of these predictive algorithms while also showing conditions where improvements can be made. In general, the forecasting capability for solar indices at low and moderate activity levels has comparably low uncertainty and virtually no bias. This performance is degraded to an extent at elevated and especially high activity levels, where the sun is more volatile. 

The best performing algorithm is for $Y_{10.7}$ whose forecasting method is the most complex of the four solar indices investigated. The algorithm for $M_{10.7}$ also has low uncertainty and low bias at the two lower solar activity levels. The forecasts for $F_{10.7}$ and $S_{10.7}$ prove to be more uncertain and with generally higher biases. Both indices had strong tendencies to over-predict at high solar activity.

The geomagnetic indices, $ap$ and $Dst$, proved to be difficult to predict even using the two diverse methods. The forecasts for $ap$ are determined by a team of forecasters with the aid of a model, and there was still a low probability of detection for geomagnetic storms. In most conditions however, there was little or no bias in the predictions. The three-day prediction window also ended up being a limitation, and results from a full six-day forecast would be intriguing. The $Dst$ algorithm performed well during G0 (or quiet) conditions but showed unusual trends with increased geomagnetic activity.

A major limitation in this study was the lack of forecasts in certain conditions. This was particularly problematic for the geomagnetic indices, and using the most extreme index value to bin forecasts was used to offset this limitation. Even with this technique, a large percentage of conditions had insufficient data to perform the uncertainty analysis. In the future, we hope to include additional forecasts to the analysis to update the results in order to cover more conditions.

This work is intended to provide the community with a performance level for future algorithm and model development in an effort to improve our capability to accurately forecast density and determine satellite trajectories.

\section{Acknowledgment}
The authors would like to acknowledge Bruce Bowman, Dave Bouwer, and Alfredo Cruz of Space Environment Technologies for access to forecasts and relevant documentation to perform this study in addition to providing important insight. This work was made possible by NASA West Virginia Space Grant Consortium, Training Grant $\#$NNX15AI01H and NASA Established Program to Stimulate Competitive Research, Grant $\#$80NSSC19M0054.

\bibstyle{AAS_Publications}   % Number the references.
\bibliography{references}   % Use references.bib to resolve the labels.

\end{document}